\documentstyle[aps,multicol,epsfig,psfig]{revtex}  
\begin{document}

\title{Two-gap model for
underdoped cuprate superconductors}
\author{A. Perali$^1$, 
C. Castellani$^1$, C. Di Castro$^1$, 
M. Grilli$^1$, E. Piegari$^2$ and A. A. Varlamov$^1$}

\address{(1) Dipartimento di Fisica, Universit\`a di Roma ``La
Sapienza'' and Istituto Nazionale Fisica della Materia,\\
Unit\`a di Roma 1, P.le A.  Moro, 2 - 00185 - Roma, Italy}
\address{(2) Dipartimento di Fisica, Universit\`a di Firenze,
L. E. Fermi, 2 - 50125 - Firenze, Italy}
\date{\today} 

\maketitle 
\medskip

\begin{abstract}
Various properties of underdoped superconducting
cuprates, including the momentum-dependent pseudogap opening,  
indicate a behavior which is neither BCS nor Bose-Einstein
condensation (BEC) like. To explain
this issue we introduce  
a two-gap model. This model assumes an anisotropic
pairing interaction among two kinds of fermions with small
and large Fermi velocities representing the quasiparticles
near the M and the nodal points of the Fermi surface
respectively. We find that a gap forms near the M points 
resulting into incoherent pairing due to strong fluctuations.
Instead the pairing near the nodal points
sets in with phase coherence at lower temperature. By tuning the 
momentum-dependent interaction,
the model allows for a continuous
evolution from a pure BCS pairing (in the overdoped and optimally doped
regime) to a mixed boson-fermion picture (in the strongly underdoped
regime).
\end{abstract}

{\small PACS numbers:74.20.De, 74.20.Mn, 71.10.-w}
\vskip 2pc 

\begin{multicols}{2}

The underdoped cuprates are characterized by a pseudogap opening 
below a strong doping $(\delta)$ dependent 
crossover temperature $T^*(\delta)$, above the
superconducting critical temperature $T_{c}(\delta)$ \cite{timusk}.
By decreasing the doping the temperature
$T^*$ increases, while the superconducting critical temperature
$T_{c}$ decreases until the insulating state is reached. 
The different behavior of $T^*$ and $T_{c}$ as doping is varied,
finds a counterpart in the different behavior
of the coherence energy scale, obtained in Andreev reflection
measurements \cite{deutscher}, and the single-particle
gap, observed both in angle-resolved photoemission (ARPES) and
in tunneling experiments \cite{timusk}. This has triggered 
a very active debate on the relevance of a 
non-BCS superconductivity and of a BCS-BEC crossover
in these materials
\cite{emerybenoy,randeria,ranninger,GIL}. In particular ARPES 
shows that  below $T^*$ the gap opens around the M points of the 
Brillouin zone  [{\em i.e.} $(\pm \pi, 0),(0,\pm \pi)$] 
suggesting that $T^*$ can be interpreted as a
mean-field-like temperature where electrons start to form local pairs
without phase coherence. However, it is also found 
\cite{norman} that 
below $T^*$ substantial portions of the Fermi surface 
remain gapless. This behavior can be described  neither 
by BCS nor by Bose-Einstein condensation (BEC) schemes. Instead
it is suggestive of strong pairing between the states around the M points
and of weak coupling near the zone diagonals.
Various other experiments \cite{panaxiang,mesot} carried below $T_c$
show a doping and temperature dependence
of the gap anisotropy and therefore are again suggestive of a
strong anisotropy in the pairing potential.

In this letter we explore a new direction (neither BCS nor BEC)
focusing on the  consequences of a strongly anisotropic
interaction. To this aim we introduce a two-gap model, where
strongly paired fermionic states 
can coexist and interplay with weakly coupled pairs in different regions
of the Fermi surface (FS). This line of thinking was partly
explored in Refs. \cite{ranninger,GIL}, where only the extreme
strong-coupling limit of one component was considered. 
In particular the view of Ref. \cite{GIL}
would allow to describe only the very underdoped
region of the cuprate phase diagram. Our approach, instead,
turns out to be sufficiently flexible to investigate with continuity 
the evolution of the bifurcation
(taking place around optimal doping in the real systems)
between a mean-field-like pairing temperature $T^*$
and the coherence superconducting temperature $T_{c}$.
We implement our model by a two-band system
with different intraband and interband
pairing interactions. One band has a large Fermi velocity
$v_F$ and a small
attraction giving rise to largely overlapped Cooper pairs with 
weak superconducting fluctuations. On the contrary, the other band 
has a small $v_F$ and a large attraction resulting into
tightly bound pairs having strong fluctuations.
At variance from the
models of mixed fermions and bosons, we keep the fermionic
nature of both the weakly and strongly bound Cooper pairs.

A possible realization of
a strongly momentum-dependent effective interaction
in underdoped cuprates has been proposed in connection to the
occurrence of a charge instability for stripe formation 
\cite{prl95,zf97}.
It was indeed suggested that the tendency to spatial charge order
(which evolves into a spin-charge stripe phase
by lowering the doping) gives rise to an instability
line. This line $T_{stripe}^c(\delta)$ starts from
a Quantum Critical Point (QCP) at $T=0$ near optimal
doping and increses by lowering the doping.
By approaching the instability line the pairing 
is mediated by the strong attractive quasi-critical stripe fluctuations,
which affect the states on the FS in a quite anisotropic way:
\begin{equation}
V_{\rm eff} ({\bf q},\omega) \approx 
\tilde{U} - \frac{V}{\kappa^{2}+
\vert {\bf q} - {\bf q}_c \vert^2
- i\gamma \omega} 
\label{fitgamlr}
\end{equation}
where $\tilde{U}$ is the residual repulsive interaction 
between the quasiparticles, $\gamma$ is a damping parameter,
and ${\bf q}_c$ is the wavevector of the stripe instability.
The crucial parameter $\kappa^2$ is a mass term proportional
to the inverse square of the correlation length of charge
order $\xi_c^{-2}$ and provides a measure of the distance
from criticality. At $T=0$, in the overdoped regime,
$\kappa^2$ is linear in the doping deviation from the
critical concentration $\kappa^2\propto (\delta-\delta_c)$.
On the other hand, in the finite-temperature region above $\delta_c$
$\kappa^2\propto T$. In the underdoped regime 
$\kappa^2$ vanishes approaching the instability line 
$T_{stripe}^c(\delta)$ and extend the singular potential to finite
temperatures. For $\kappa^2 \approx 0$, 
the fermionic states around the M points are such that
${\bf k}_F -{\bf k_F}' \sim {\bf q_c}$ and interact strongly.
These are the so-called ``hot spots''\cite{notafm}, 
they have a low dispersion, and possibly 
form tightly bound local pairs giving rise to
the pseudogaps below $T^* \gtrsim T_{stripe}^c(\delta)$. 
On the other hand, ``cold'' states
in the arcs of FS around the zone diagonals $\Gamma$-$Y$  
or $\Gamma$-$X$ (nodal points) have larger dispersions and
interact more weakly since $V_{\rm eff}$ is now cut-off by ${\bf q}_c$.
In the underdoped regime $\kappa^2 \approx 0$ at higher and higher
temperatures by lowering the doping and $V_{\rm eff}$ has a more
dramatic effect. On the contrary, in going to the optimum 
and the overdoped region
$V_{\rm eff}$ is cut-off first by the temperature and then by the
doping itself. All the states then interact more isotropically.

{\em --- The two-gap model.} 
Irrespectively of the origin of the anisotropy, in the presence of 
a strong momentum dependence both of ($i$) the effective pairing 
interaction and ($ii$) the Fermi velocity, 
we must allow for enough freedom
of the pairing and of its fluctuations in order to capture the relevant 
physical effects of the anisotropy. 
Following the above discussion we introduce a simple two-band model
for the cuprates.
We describe the quasiparticle arcs of FS 
about the nodal points by a free electron
band  (labelled below by the index 1)
with a large Fermi velocity $v_{F1}=k_{F1}/m_1$ and the 
hot states about the M points with a second free electron band, 
displaced in momentum and in energy from the first, with a small
$v_{F2}=k_{F2}/m_2$ (see Fig. 1) \cite{bianconi}.\vspace{-0.5 truecm}
\begin{figure}
\begin{center}
\psfig{figure=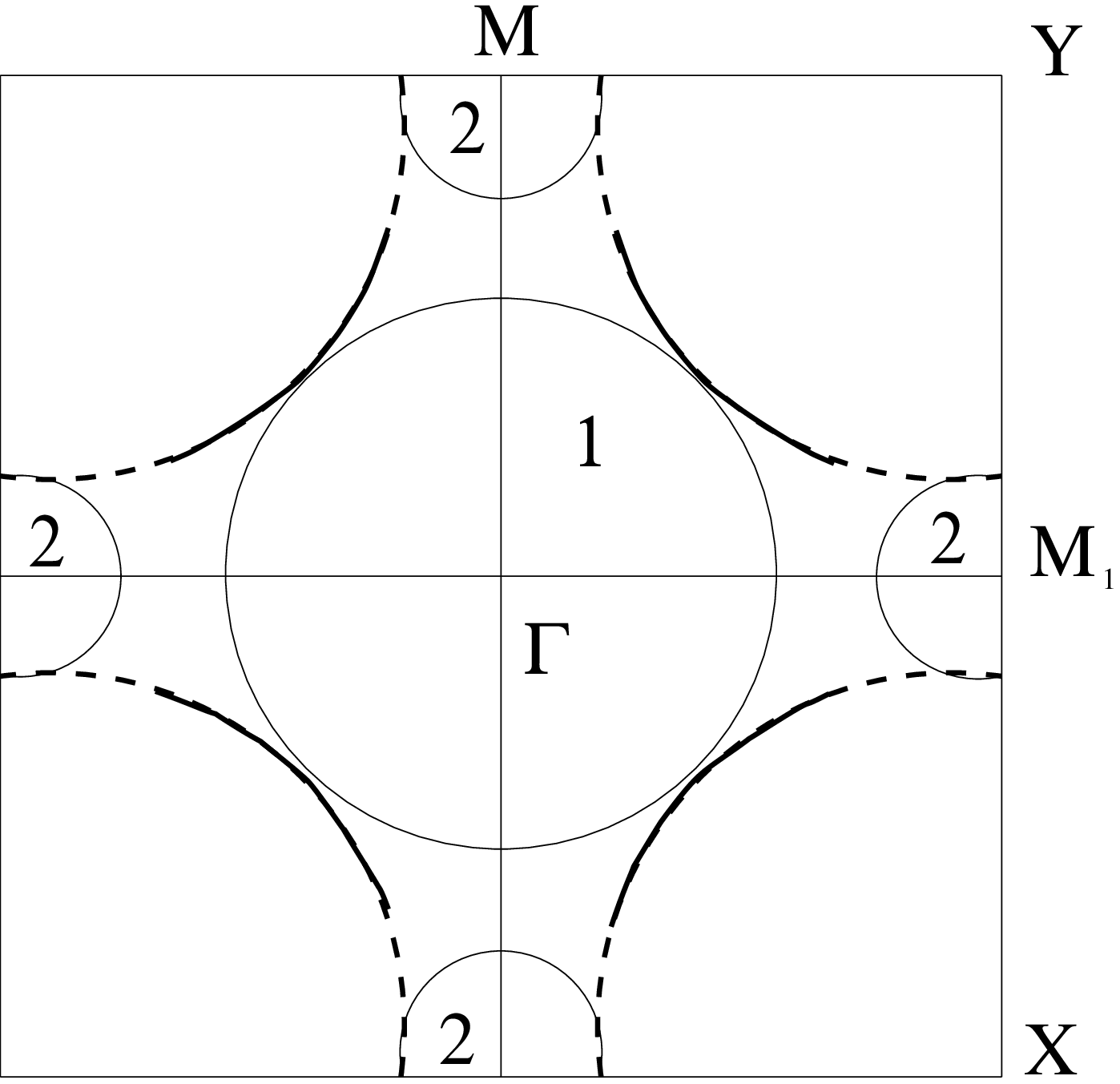,width=4cm,height=4cm,angle=-90}
\label{fsmodel}
\end{center}
\end{figure}
\vspace{-0.5 truecm}
{\small Fig. 1: Sketch of the Fermi surface of 
underdoped cuprates with quasiparticle arcs
(thick solid line) and patches of quasi-localized states 
(thick dotted line) and the Fermi
surface of the two-band spectrum (thin solid line).}
\vspace{0.5 truecm}

The assumed electronic spectrum consists therefore of two different 
free electron dispersions, 
$\varepsilon_1({\bf p})=\mid {\bf p}\mid^2/2m_1$ and 
$\varepsilon_2({\bf p})=\mid {\bf p}- {\bf p}_0\mid ^2/2m_2+\varepsilon_0$,
displaced by a momentum ${\bf p}_0$ 
and by an energy $\varepsilon_0\sim E_{F1}$ introduced to
allow the chemical potential to cross
both bands: $E_{F1}=\varepsilon_0+E_{F2}$.
To connect our two-band structure with the single-band 
dispersion of the
cuprates, we choose ${\bf p}_0=(\pm \pi,0),(0,\pm \pi)$. 

This choice gives rise to two
branches for the band 2 along the $x$ and $y$ directions. 
Moreover,  we only consider Cooper pairs of zero total 
momentum formed by time-reversed momentum eigenstates. Therefore
the 2-2 pairs are always formed by $({\bf k},-{\bf k})$ states on 
portions of the band 2 symmetrically located on opposite sides 
with respect to the $\Gamma\equiv (0,0)$ point. 
If the pairs have a $s$-wave
symmetry, the branches along $x$ and $y$ of the band 2
are equivalent and just one can be considered. On the other hand,
in the case of $d$-wave pairing, the pairs in the branch along $x$ 
have  a different phase from the pairs in the branch along $y$
and both branches have to be treated. Since in this paper
our main interest is the interplay between strongly and weakly coupled
pairs irrespectively of their symmetry, for simplicity we
consider the $s$-wave problem .
The model Hamiltonian for pairing in the two-band system is
taken to be

\begin{equation}
H=\!
\sum_{k\sigma i} \varepsilon_{ki}
n_{k\sigma i}+
\!\!\!\sum_{kk^{\prime}p i j} \!\!\!V_{ij}(k,k^{\prime})
c^+_{k^{\prime}+p\uparrow j}c^+_{-k^{\prime}\downarrow j}
c_{-k\downarrow i}c_{k+p\uparrow i}
\label{hamiltonian}
\end{equation}
where $i$ and $j$ run over the band 
indices 1 and 2 and $\sigma$ is the spin index.
The interaction is approximated by a BCS-like attraction given by
\begin{equation}
V_{ij}(k,k^{\prime})=-g_{ij}
\Theta(\omega_0-\mid\xi_{i}(k)\mid)
\Theta(\omega_0-\mid\xi_{j}(k^{\prime})\mid)
\label{vbcs}
\end{equation}
with an energy cutoff  $\omega_0$.
The strongly $q$-dependent effective interaction in the particle-particle
channel $V_{\rm eff}({\bf q}={\bf k}-{\bf k}^{\prime})$
of the original single-band system is accounted for by the
$2\times 2$ scattering matrix $\hat g$. The matrix elements $g_{ij}$
couple the electrons within the
same band ($g_{11}$ and $g_{22}$) and between different bands 
($g_{12}=g_{21}$). 
The self-consistency equation for the superconducting fluctuation
propagator  \cite{Varlamov} in the matrix form is given by
$\hat{L}=\hat{g}+\hat{g}\hat{\Pi}\hat{L}$, where
the  particle-particle bubble operator for the two-band spectrum
has a diagonal $2\times 2$ matrix form with elements
$\Pi_{11}(q)$ and $\Pi_{22}(q)$ \cite{notapi12}. The resulting 
fluctuation propagator is given by
\begin{equation}
\hat{L}(q)=
\left(\begin{array}{cc}
\tilde{g}_{11}-\Pi_{11}(q) &             \tilde{g}_{12}\\
\tilde{g}_{12}             & \tilde{g}_{22}-\Pi_{22}(q)\\
\end{array}\right)^{-1};
\label{propmatrix}
\end{equation} 
where we have defined $\tilde{g}_{ij}\equiv (\hat{g}^{-1})_{ij}$.
It turns out useful 
to define the temperatures $T_{c1}^0$ and $T_{c2}^0$ as
$\tilde{g}_{11}-\Pi_{11}(0,T)=\tilde{g}_{11}
-\rho_1\ln\frac{\omega_0}{T}\equiv \rho_1 \ln \frac{T}{T_{c1}^0}$,
$\tilde{g}_{22}-\Pi_{22}(0,T)=\tilde{g}_{22}
-\rho_2\ln\frac{\omega_0}{T}\equiv \rho_2 \ln \frac{T}{T_{c2}^0}$,
where $\rho_i=m_i/(2\pi)$ 
is the density of states of the $i$-th band.
In the underdoped regime, 
to emulate the hot and cold points related for instance to the
system near a stripe instability, we assume the following relations
between the $g_{ij}$ elements: $g_{22} \simeq V/\kappa^2
>>g_{11} \simeq V/|{\bf q}_c|^2 \simeq g_{12} $.
Then one has 
$
\tilde{g}_{11}\simeq 1/g_{11}, \,\,
\tilde{g}_{22}\simeq 1/g_{22}, \,\,
\tilde{g}_{12}\simeq -g_{12}/(g_{11}g_{22}).
$
In this limit $T_{c1}^0$ and $T_{c2}^0$ (with $T_{c2}^0 \gg T_{c1}^0$)
assume the value of
 the two BCS critical temperatures for the two decoupled bands. 
The mean-field BCS superconducting critical temperature 
for the coupled system $T_c^0$ is defined by
the equation $\mbox{det}\hat{L}^{-1}({\bf q}=0, T_c^{0})=0$.
We obtain $T_c^{0}>T_{c2}^{0}$ given by 
\begin{equation}
T_c^{0}=\sqrt{T_{c1}^{0}T_{c2}^{0}}\exp\left[{\frac{1}{2}
\sqrt{\ln^2\left(\frac{T_{c2}^{0}}{T_{c1}^{0}}\right)+
\frac{4\tilde{g}_{12}^2}{\rho_1\rho_2}}}\right].
\label{tcmf}
\end{equation} 

{\em --- The Ginzburg-Landau approach.} 
The role of fluctuations can be investigated within a
standard Ginzburg-Landau (GL) scheme, 
when both $\rho_2g_{22}\omega_0< E_{F2}$ and $\omega_0< E_{F2}$.
Under these conditions the chemical potential is not affected 
significantly by pairing. Moreover, in order to remain within 
the GL approach, we will assume
that fluctuations from the BCS result are not too strong.
The relevance of the space fluctuations of the order parameter
is assessed by  the gradient term coefficient
$\eta$, which provides the momentum dependence of the fluctuation 
propagator in Eq.(\ref{propmatrix}). 
In particular the expansion of the particle-particle 
bubbles, in terms of $q$, reads 
$\Pi_{11}(q)\simeq\Pi_{11}(0)-\rho_1\eta_1 q^2$ and 
$\Pi_{22}(q)\simeq\Pi_{22}(0)-\rho_2\eta_2 q^2$. 
Here $\eta_i(i=1,2)$ is the gradient term coefficient of the
$i$-th band which, in 2D and for a free electron band, 
is given by $\eta_i=(7\zeta(3)/32\pi^2)v_{Fi}^2/T^2$,
with  $\eta_1\gg\eta_2$ \cite{notaeta}.
In the absence of the interband coupling $g_{12}$,
$\eta_1$ provides the (large) gradient term coefficient 
corresponding to (the weak) superconducting fluctuations for the 
band with a large $v_{F1}$, while (the small)
$\eta_2$ corresponds to (strong) superconducting
fluctuations for the band 2.
For the coupled system near $T_c^0$, 
the  coefficient $\eta$ in terms of $\eta_1$ and $\eta_2$
is obtained by evaluating $({\hat L}^{-1})_{ij}\propto
\left( \epsilon+\eta q^2 \right)$ in terms of the relative
temperature deviation $\epsilon \equiv (T-T_c^0)/T_c^0$.
We get the expression 
\begin{equation}
\eta=\frac{\rho_1(\tilde{g}_{22}-\Pi_{22})\eta_1+
\rho_2(\tilde{g}_{11}-\Pi_{11})\eta_2}
{-T(\tilde{g}_{22}-\Pi_{22})d\Pi_{11}/dT
-T(\tilde{g}_{11}-\Pi_{11})d\Pi_{22}/dT},
\label{etalong}
\end{equation}
where all the bubbles are evaluated at $q=0$. Using 
the definitions for $T_{c1,2}^0$ and the condition
$\mbox{det}\hat{L}^{-1}({\bf q}=0, T_c^{0})=0$, the coefficient
$\eta$ can be explicitly written as
\begin{equation}
\eta=\alpha_1\eta_1+\alpha_2\eta_2,\; {\rm with}\;\;
\frac{\alpha_1}{\alpha_2}=\frac{\tilde{g}^2_{12}}
{\rho_1\rho_2\ln^2(T_c^{0}/T_{c1}^{0})}, 
\label{etaeff}
\end{equation}
and $\alpha_1+\alpha_2=1$.
The presence of a fraction of electrons with a large 
$\eta_1$ increases the stiffness 
 of the whole electronic system ({\em i.e.} increases $\eta$
with respect to $\eta_2$). 
However when the mean-field critical temperature $T_c^{0}$ is much 
larger than $T_{c1}^{0}$ the correction to $\eta_2$ due to the
interband coupling is small. At the same time the Ginzburg number 
is large implying a sizable
mass correction $\delta\epsilon (T)$ to the ``mass'' 
$\epsilon (T)$ of the bare propagator $\hat{L}(q)$. 
The renormalized critical temperature $T_c^{r}$, 
given by the equation
$\epsilon(T_c^r)+\delta\epsilon (T_c^r)=0$,
is lower than $T_c^{0}$ \cite{18proceed}.
By evaluating the renormalized gradient term coefficient 
$\eta^r$ in the presence of the mass correction,
we find that this is still given by Eq. (\ref{etaeff})
with $T_c^0$ replaced by $T_c^r$. Therefore 
$\eta^r=\eta(T_c^r)$ is greater than
$\eta (T_c^{0})$.
This result indicates that the mass renormalizations of the
fluctuation propagator tends to lower $T_c$ and, at the same time,
increases the gradient term coefficient $\eta$ 
by increasing the coupling to $\eta_1$. As a consequence
the effective Ginzburg number is reduced and the system is stabilized with 
respect to fluctuations allowing for a coherent superconducting phase
even in the limit $\eta_2 \to 0$.
Within the GL approach we associate the temperature
 $T_{c}^{0}\sim T_{c2}^{0}$ to
the crossover temperature $T^*$ and $T_c^r$ to the superconducting 
critical temperature $T_{c}$ of the whole system. 
In the region $T_c<T<T^*$ the pseudogap is formed in band 2.
Superconducting fluctuations only affect band 2 while
are immaterial for band 1, where the Fermi surface is mantained
until phase coherence sets in.
Within the Stripe-QCP scenario the coupling 
$g_{22}$ is related to the singular part 
of the effective interaction mediated by the stripe fluctuations.
$g_{22}$ is the most doping dependent coupling and attains 
its largest value in the underdoped regime, where
$\kappa^2$ vanishes at higher temperatures.
The regular parts of the interaction 
$g_{11}$ and $g_{12}$, being cut-off by ${\bf q_c}$, 
 are instead only weakly doping-dependent.
In the region of validity of the GL approach,
the explicit calculations show that $r(\delta)\equiv
{(T^0_c-T^r_c)}/{T^0_c}
\simeq {(T^*-T_c)}/{T^*}$ is increasing by increasing 
$g_{22}$, {\em i.e.}, by decreasing doping. For small values of $r$
we find that both $T^*$ and $T_c$ increase. 
This regime corresponds to the overdoped and optimally doped region.
Specifically, above optimum doping, $g_{22}$ and $g_{11}$ become comparable
and the two lines merge together. 
For $r\sim 0.25\div 0.5$, $T_c$ is instead decreasing while 
$T^*$ is always increasing by decreasing doping. The large values
of $r$, which are attained in the underdoped region show that we
are reaching the limit of validity of our GL approach. 
We think, however, that the behavior of the
bifurcation between $T^*$ and $T_c$ represents correctly 
the physics of the pseudogap phase, while a
quantitative description would require a more sophisticated approach
like a RG analysis.

{\em --- The strong-coupling limit.} 
In the very low doping regime, where $T^*$ has increased strongly,
the value of $g_{22}$ can be so large 
to drive the system in a strong coupling regime for the
fermions in band 2 ($\rho_2 g_{22}\omega_0> E_{F2}$).
In this case, taking $\omega_0 > E_{F2}$ the chemical potential 
is pulled below the bottom of the band 2. 
In this limit of tightly bound 2-2 pairs,
the propagator of the superconducting
fluctuations in band 2, {\em i.e.} 
$L_{22}(q)$, assumes the form of a single pole
for a bosonic particle. Since  $ E_{F1}$ is still
the largest energy scale  in the problem,
the fermionic character of the particles in band 1
is preserved.
The critical temperature of the system is again obtained by the vanishing 
of $\det {\hat L}^{-1}(q=0)$ where, 
however, the chemical potential is now self-consistently evaluated 
including the selfenergy corrections to the Green function in band 2
and the fermions left in band 1. One gets  
$\Pi_{22}=\rho_2\omega_0/\vert \mu_2 \vert$ and 
\begin{equation}
T_c^0=T_{c1}^0 \exp\left[\frac{\tilde{g}_{12}^2}{\rho_1\rho_2\omega_0}
\frac{|\mu_2||\mu_B|}{(|\mu_2|-|\mu_B|)}\right]
\label{eq10}
\end{equation}
where $\mu_2$ is the chemical potential measured with respect to the 
bottom of the band 2 and $|\mu_B|=\rho_2g_{22}\omega_0$
represents the bound-state energy.
The calculation of $\Pi_{22}$ at small $q$ leads to a finite value of 
$\eta_2=1/(8m_2|\mu_2|)$, while the small-$q$ expansion of
$\det {\hat L}^{-1}$ provides the new $\eta$ coefficient
\cite{notaetasc}
\begin{equation}
\eta = \eta_1+ 
\rho_1\rho_2\omega_0\frac{g_{22}^2}{|\mu_2|}
\ln^2(T_c^0/T_{c1}^0)\eta_2
\label{etasc}
\end{equation}
In this strong-coupling case
most of the non-mean-field effect has been taken into account
by the formation of the bound state occurring at a very high
temperature of the order $\rho_2g_{22}\omega_0$, which provides
the new $T^*$ in this regime. 
$\eta \sim \eta_1$ stays sizable and
the fluctuations will not strongly further
reduce $T_c^r$ with respect to 
$T_c^0$: $T_c\simeq T_c^r \simeq T_c^0$.
In this low-doping regime $\frac{T^*-T_c}{T^*}$ approaches its largest
values before $T_c$ vanishes.

The strong coupling limit of
our model shares some similarities as well as some important differences
with  phenomenological models of interacting bosons and fermions 
\cite{ranninger,GIL}. In particular, in the model of Ref. \cite{GIL}
pairs of non interacting electrons are 
scattered from the Brillouin zone near the nodal points
into dispersionless boson
states, localized about the M points  at an energy $\varepsilon_B$.
The correspondence with
our model can be seen by noticing that the tightly bound bosonic
states correspond to $\rho_2g_{22}\omega_0\gg E_{F2}$ 
and $g_{11}=0$. The tightly bound
dispersionless bosonic states are fully incoherent
($\eta_2=0$), while the fermionic
states are unpaired as long as bosons and fermions are independent. 
The fermion-boson coupling
$g_{12}$ effectively introduces an $e$-$e$ coupling of the order of
$\rho_1 \rho_2 \omega_0
g_{12}^2/\varepsilon_B$ and drives the system superconducting.
In this particular limiting case of our model 
we recover the results of Ref.
\cite{GIL} with an
explicit expression for the bosonic level
$\varepsilon_B=|\mu|-|\mu_B|$, while in Ref. \onlinecite{GIL}
it is a free phenomenological parameter. 

{\em --- Conclusions.}
In our paper we have analyzed the pairing properties 
of the underdoped
cuprates in terms of an effective two-gap model, motivated by 
the strong anisotropy of the
band dispersions and of the effective 
pairing interaction. 
The crucial schematization was based on the introduction
of two bands weakly coupled in order to preserve a substantial
distinction between the superconducting order parameter in 
different regions of the momentum space. This has to be 
contrasted with
more standard approaches
producing one single gap (even with complicated momentum
dependence) and a common fluctuating order parameter all over the FS.
Our approach allows for different 
fluctuation regimes for pairs in different 
${\bf k}$ regions. According to our analysis, 
the strongly bound pairs forming at an high temperature $T^*$
can experience large fluctuations until the system is stabilized by
the coupling with less bound BCS-like states,
leading to a coherent superconducting state at $T_c<T^*$. 
$T_c$ and $T^*$ merge around or above optimum doping,
where $g_{22}$, according to our stripe scenario,
becomes of the order of $g_{11}$. 
Our model shares
similarities with the fermion-boson models for cuprates
\cite{ranninger,GIL} to which it reduces 
in the  strong-coupling limit for $g_{22}\gg g_{11}\simeq 0$. 
The two-gap model considered here applies 
to a much wider doping region
and is more suitable to describe
the crossover to the optimal and over-doped regime,
where no preformed bound states are present and the superconducting 
transition is quite similar to a standard BCS transition.

{\em Acknowledgments.} We acknowledge S. Caprara
for helpful discussions.

\end{multicols}

\begin{thebibliography}{99}
\bibitem{notapi12} Since the model only considers pairing between
	time-reversed states with zero total momentum, fermionic 
	particle-particle bubbles involving 
	one fermion on band 1 and 
	one fermion on band 2 never appear in the ladder resummation
	for the propagator $\hat L(q)$.
\bibitem{18proceed}  In the explicit calculation we assume that the 
	largest contribution to $\delta\epsilon (T)$ comes from the first 
	order selfenergy corrections to the bare Green functions of the 
	$p$-$p$ bubbles of Eq.(\ref{propmatrix}). In these corrections we 
	cutoff the $2D$ logarithmic divergencies by a small interlayer 
	coupling.
\bibitem{notaetasc} Neglecting the
	temperature dependence of $\Pi_{22}$ in Eq. (\ref{etalong}) and
	using Eq. (\ref{eq10}), one can show that, despite the different 
	physical situation, this strong-coupling form of $\eta$
	assumes the same expression as in Eq. (\ref{etaeff}).
\end{thebibliography}
\end{document}